\documentclass{aa}
\usepackage{graphicx}
\usepackage{txfonts}
\usepackage{tabularx}
\usepackage{array}
\usepackage{amssymb}
\newcommand{\Msun}{\ensuremath{\mathrm{M}_\odot}}

\newcommand{\lsim}{\mathrel{\hbox{\rlap{\lower.55ex \hbox {$\sim$}}
 \kern-.3em \raise.4ex \hbox{$<$}}}}
\newcommand{\gsim}{\mathrel{\hbox{\rlap{\lower.55ex \hbox {$\sim$}}
 \kern-.3em \raise.4ex \hbox{$>$}}}}

\def\mso{\,\mathrm{M}_\odot}

\def\simle{\mathrel{\hbox{\rlap{\hbox{\lower4pt\hbox{$\sim$}}}\hbox{$<$}}}}
\def\simgr{\mathrel{\hbox{\rlap{\hbox{\lower4pt\hbox{$\sim$}}}\hbox{$>$}}}}

\begin{document}
   \title{Evolution of rapidly rotating metal-poor massive stars towards gamma-ray bursts}


   \author{S.-C. Yoon
         \inst{1}
          \and
          N. Langer
          \inst{2}
          }

   \offprints{S.-C. Yoon}

     \institute{Astronomical Institute "Anton Pannekoek", University of Amsterdam, 
               Kruislaan 403, 1098 SJ, Amsterdam, The Netherlands 
         \and
              Astronomical Institute, Utrecht University,
              Princetonplein 5, NL-3584 CC, Utrecht, The Netherlands
              \\
              \email{scyoon@science.uva.nl; n.langer@astro.uu.nl}
             }

   \date{Received ; accepted }

   \abstract{
              Recent models of rotating massive stars including magnetic fields prove 
             it difficult for the cores of single stars 
             to retain enough angular momentum to produce a collapsar and gamma-ray burst.
              At low metallicity, even very massive stars may retain a massive 
              hydrogen envelope due to the weakness of the stellar winds, 
              posing an additional obstacle to the collapsar model.
              Here, we consider the evolution of massive, magnetic stars where rapid rotation 
             induces almost chemically homogeneous evolution.
             We find that in this case, the requirements of the collapsar model
             are rather easily fulfilled if the metallicity sufficiently small:
             1) Rapidly rotating helium stars are formed without the need to
             remove the hydrogen envelope, avoiding mass-loss induced spin-down. 
             2) Angular momentum transport from the helium core to hydrogen envelope 
             by magnetic torques is insignificant. We demonstrate this by calculating
             evolutionary models of massive stars with various metallicities, 
             and derive an upper metallicity limit for this scenario based on 
             currently proposed mass loss rates. Our models also suggest the existence
             of a lower CO-core mass limit of about $10\mso$ --- which relates to an
             initial mass of only about $20\mso$ within our scenario --- for GRB production.   
             We argue that the relative importance of the considered
             GRB progenitor channel, compared to any channel related to  binary stars, 
             may increase with decreasing metallicity, and
             that this channel might be the major path to GRBs from first stars.
             
   }
   \titlerunning{Evolution of metal poor stars towards GRBs}
   \authorrunning{Yoon \& Langer}
   \maketitle
%

\section{Introduction}

As long gamma-ray bursts (GRBs) are observable, in principle, 
from redshifts as high as 20 (e.g. Wijers et al.~\cite{Wijers98}),
their study may yield break-throughs in our understanding of the
early universe. Their association with the deaths of massive stars
(e.g. Hjorth et al.~\cite{Hjorth03}; Stanek et al.~\cite{Stanek03})
renders them as tracer of the cosmic history of star formation
(see Bromm \& Larson~\cite{Bromm04} for a review).
Their possible use as standard candles (e.g. Friedman \& Bloom~\cite{Friedman05}; Lamb et al.~\cite{Lamb05})  
may turn the into the farthest reaching cosmic yardstick.
However, our present lack of understanding the progenitor evolution of
long GRBs hampers their deeper understanding and quantitative exploitation
for cosmology.

According to the currently favoured collapsar scenario, 
relativistic jets are formed
when rapidly rotating iron cores of massive helium stars collapse
to form  black holes and accretion disks around them (Woosley~\cite{Woosley93}).  
However, recent stellar evolution models which include 
angular momentum transport from the core to the hydrogen envelope by magnetic torques
(Spruit~\cite{Spruit02})
indicate the most single stars end up with too slowly rotating cores.
Especially during the giant stage, where strong differential rotation persists
between helium core and hydrogen envelope, angular momentum is efficiently removed from
the core (Maeder \& Meynet~\cite{Maeder03}; Heger et al.~\cite{Heger05};
Petrovic et al.~\cite{Petrovic05}; see also below).
Although some non-magnetic stellar models are found to fulfill the necessary conditions for collapsar 
(Heger et al.~\cite{Heger00b}; Petrovic et al.~\cite{Petrovic05}; Hirschi et al.~\cite{Hirschi05}), 
they predict a much too high GRB vs. Type~Ibc supernova ratio (e.g., van Putten~\cite{Putten04}). 
The inclusion of magnetic torques is further required to reproduce the observed low rotation
rates of young neutron stars (Heger et al.~\cite{Heger05}) and white dwarfs 
(Suijs et al.~\cite{Suijs05}).

The formation of collapsars is further complicated by considering mass loss. For rather
large mass loss rates, a rotating star is efficiently spun-down even without magnetic effects
(Langer~\cite{Langer98}). This is in particular true also for massive helium stars, which, 
in our Galaxy, appear as Wolf-Rayet star with wind dominated spectra: 
any long-lasting Wolf-Rayet phase at solar metallicity will likely lead to smaller spins
than required by the collapsar model. However, Wolf-Rayet mass loss rates may be
smaller for smaller metallicity (Nugis \& Lamers~\cite{Nugis00} (NL00);
Crowther et al.~\cite{Crowther02}; Vink \& de Koter~\cite{Vink05}).
The problem then is that the winds of potential Wolf-Rayet star progenitors
also weaken for lower metallicity (Kudritzki et al.~\cite{Kudritzki87}; Kudritzki~\cite{Kudritzki02}),
with the consequence that the Wolf-Rayet stage may actually never be reached
(Heger et al.~\cite{Heger03}; Marigo et al.~\cite{Marigo03};  Meynet \& Maeder~\cite{Meynet02}).

Here, we explore an evolutionary channel which can avoid both, the problem of magnetic torques,
and the mass loss problem, at low metallicity: chemically homogeneous evolution.
Maeder~(\cite{Maeder87}) found that if rotationally induced chemical mixing in massive
main sequence stars occurs faster than the built-up of chemical gradients due to nuclear fusion,
the chemical gradients may remain small throughout core hydrogen burning, transforming
the main sequence star smoothly into a helium star of the same mass. Even though the physics of
rotationally induced mixing has been refined since then, quasi-chemically homogeneous evolution
is still found if very rapid rotation is adopted (Langer~\cite{Langer92}; 
Heger \& Langer~\cite{Heger00a}; Maeder \& Meynet~\cite{Maeder00}), and remains possible
also when magnetic torques are included (see below).

In quasi-chemically homogeneously evolving models, the helium star is born without, or with
only a small hydrogen-rich envelope. Magnetic torques can therefore remove only little
angular momentum from the helium star. And at low enough metallicity, the mechanical mass loss
induced spin-down may be avoided due to the weakness of the helium star winds. 
After we explain our numerical methods and physical assumptions in Sect.~\ref{sect:methods},
we present models of the homogeneous evolution of rapidly rotating massive stars at low metallicity
with the effect of magnetic torques included in Sect.~\ref{sect:results}.
We discuss the potential implications of our results for an understanding of
GRB progenitors in in Sect.~\ref{sect:discussion}.

\section{Methods and physical assumptions}\label{sect:methods}

Stellar models are calculated with a hydrodynamic stellar evolution code
(cf. Petrovic et al.~\cite{Petrovic05}, and references therein), 
which includes the effect of the centrifugal force on the stellar structure,
chemical mixing and transport of angular momentum due to rotationally 
induced hydrodynamic instabilities,
and the transport of angular momentum due to magnetic torque
(Spruit~\cite{Spruit02}; Maeder \& Meynet~\cite{Maeder05}). 

\begin{figure}
\center
\resizebox{0.9\hsize}{!}{\includegraphics{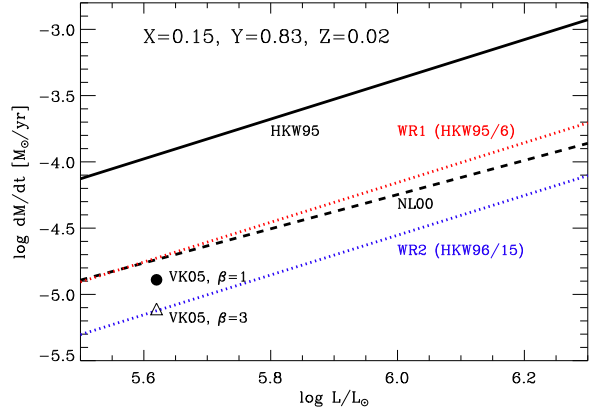}}
\caption{Wolf-Rayet wind mass loss rates as a function of the stellar luminosity
for a surface composition of $X=0.15$, $Y=0.83$, and $Z=0.02$. 
The mass loss rates by Hamann et al. (\cite{Hamann95}; HKW95) and
Nugis \& Lamers (\cite{Nugis00}; NL00) are given by
the solid and dashed lines, respectively.
The upper and lower dotted lines denote the HKW95 rates divided by a factor of 6
and 15 respectively, as assumed in this study.
The filled circle and the open triangle are 
the mass loss rates given by Vink \& de Koter (\cite{Vink05})
for WN stars, with two different assumed values of $\beta$, 
which determines the wind velocity law 
(see Vink \& de Koter~\cite{Vink05} for the exact
definition of $\beta$). 
}\label{fig:wind}
\end{figure}

Stellar wind mass loss for O stars is calculated according to Kudritzki et al. (\cite{Kudritzki89}).
Wind mass loss rates of helium rich stars (WR stars) are currently very uncertain. 
Recent WR star wind models, which include the effect of wind inhomogeneities (clumping), 
give significantly lower values than earlier estimates, e.g. by Hamann et al. (\cite{Hamann95}; 
hereafter HWK95), as shown in Fig.~\ref{fig:wind}.
Here, we use two different Wolf-Rayet loss rates (WR1 and WR2), as follows:
\begin{eqnarray}
\mathrm{WR1:}~~\log\left(\dot{M} / [\mathrm{M_\odot~yr^{-1}}]\right) & = & 
  -12.73 + 1.5\log L/\mathrm{L_\odot} - 2.85 X_\mathrm{s} \nonumber \\ 
  & & + 0.5\log(Z/Z_\odot)~, 
\end{eqnarray}
and
\begin{eqnarray}
\mathrm{WR2:}~~\log\left(\dot{M} / [\mathrm{M_\odot~yr^{-1}}]\right)  & = &
  -13.13 + 1.5\log L/\mathrm{L_\odot} - 2.85 X_\mathrm{s} \nonumber \\ 
  & &+ 0.86\log(Z/Z_\odot)~,
\end{eqnarray}
where $X_\mathrm{s}$ denotes the surface hydrogen abundance. 
Both rates correspond to HWK95, but reduced by factors 6 and 15, 
and the metallicity dependence is
$\dot{M} \propto Z^{0.5}$ and $Z^{0.86}$ for WR1 and WR2, respectively.
respectively (see Fig.~\ref{fig:wind}). 
The former is commonly used in stellar models (e.g. Heger et al.~\cite{Heger03};
Hirschi et al. ~\cite{Hirschi05}), 
while the latter is based on the recent WN stars wind models by Vink \& de Koter (\cite{Vink05}). 

The enhancement of stellar mass loss when the star approaches
the $\Omega$-limit is considered following Langer (\cite{Langer97}):
\begin{equation}
\dot{M} = \dot{M}(V_\mathrm{rot}=0)\left(\frac{1}{1-V/V_\mathrm{crit}}\right)^{0.43}~, 
\end{equation}
where
\begin{equation}
V_\mathrm{crit} = \frac{GM}{R}(1-\Gamma);~ \Gamma = \frac{\kappa L}{4\pi cGM}~.
\end{equation}
We include the effect of the enrichment of CNO elements at the 
stellar surface on the opacity, and its influence on the Eddington factor $\Gamma$. 

Initial rotation rates of our models are chosen such that rotational velocity at the 
equatorial surface
has a specified fraction of the Keplerian value: $f_\mathrm{K} = V_\mathrm{init}/V_\mathrm{Kepler}$. 
We compute models for $Z=10^{-5}$, $Z=10^{-3}$, and one model with $z=0.02$ (see Table~\ref{tab1}).
The chemical composition 
is assumed to have the same proportionality with those of solar metallicity.
More details about the numerical code can be found in Petrovic et al.~(\cite{Petrovic05})
and references therein.

\section{Results}\label{sect:results}

\begin{table*}[th]
\begin{center}
\caption{Basic model properties: Sequence number,  
$M_\mathrm{init}$: initial mass, $Z$: metallicity, 
$\dot M_{\mathrm{WR}}$: option for the WR wind mass loss rate, 
$f_\mathrm{K}$: initial rotational velocity in units of the Keplerian value, 
$V_\mathrm{init}/V_\mathrm{crit}$: the ratio of the initial rotational velocity to the critical velocity at the equator (Eq.~4), 
$Y_\mathrm{s, H-end}$: surface helium abundance at core hydrogen exhaustion, 
$M_\mathrm{final}$: final mass, 
$M_\mathrm{CO}$: CO core mass at core carbon exhaustion,  
$J_\mathrm{init}$: initial angular momentum, 
$J_\mathrm{final}$: final angular momentum, 
$j_\mathrm{3M_\odot}$: mean specific angular momentum in the innermost $3\mso$,
$X_\mathrm{N, max}$:
maximum nitrogen surface mass fraction during the evolution.
}\label{tab1}
\vspace{0.1cm}
\begin{tabularx}{\linewidth}{ l c c %
c c c  >{\centering\arraybackslash}X  >{\centering\arraybackslash}X %
>{\centering\arraybackslash}X  >{\centering\arraybackslash}X  >{\centering\arraybackslash}X  %
>{\centering\arraybackslash}X >{\centering\arraybackslash}X >{\centering\arraybackslash}X} 
\hline
\hline
No. & $M_\mathrm{init}$ & $Z$ & $\dot M_{\mathrm{WR}}$  &$V_\mathrm{init}$ & $f_\mathrm{K}$ & $V_\mathrm{init}/V_\mathrm{crit}$   & $Y_\mathrm{s, H-end}$ & $M_\mathrm{final}$ & $M_\mathrm{CO}$ &  %
$J_\mathrm{init}$ & $J_\mathrm{final}$ & $j_\mathrm{3M_\odot}$ & $X_\mathrm{N, max}$ \\
    & [$\mathrm{M_\odot}$] &  &   & [$\mathrm{km~s^{-1}}$]  &  &  &  &  [$\mathrm{M_\odot}$] &  [$\mathrm{M_\odot}$] &  
[$10^{51}$ erg s] & [$10^{51}$ erg s] & [$10^{15}$ $\mathrm{cm^2~s^{-1}}$] &   \\ 
\hline
A1  & 20 & $10^{-5}$ & WR1  &479 & 0.5 & 0.56   & 0.94   & 16.40  & 9.48  & 46.92  &  3.79 & 2.50 & $2.6 10^{-6}$ \\
A2  & 30 & $10^{-5}$ & WR1  &501 & 0.5 & 0.58   & 0.94   & 24.13  & 18.80  & 98.80  &  8.40 & 10.57 & $1.3 10^{-2}$ \\
A3  & 40 & $10^{-5}$ & WR1  &230 & 0.2 & 0.25   & 0.27   & 39.84  & 7.08   & 73.58  & 65.36 & 0.90  & $4.2 10^{-6}$\\
A4  & 40 & $10^{-5}$ & WR1  &555 & 0.5 & 0.60   & 0.94   & 31.77  & 25.23  & 166.21 & 12.57 & 12.57 & $1.4 10^{-2}$ \\
A5  & 50 & $10^{-5}$ & WR1  &583 & 0.5 & 0.64   & 0.94   & 39.21  & 33.68  & 247.54 & 19.03 & 13.75 & $3.2 10^{-2}$ \\
A6  & 60 & $10^{-5}$ & WR1  &605 & 0.5 & 0.63   & 0.94   & 46.50  & 40.25  & 341.70 & 23.55 & 13.41 & $2.8 10^{-2}$ \\
\hline
B1  & 40 & $0.001 $  & WR1  &479 & 0.5 & 0.63  & 0.90   & 13.52  & 10.70  & 183.97 & 0.21  & 0.82  & $6.7 10^{-4} $ \\
B2$^*$  & 40 & $0.02 $   & WR1  &408 & 0.5 & 0.69  & 0.69   & 20.25  &  --    & 166.40 & 1.505 &  3.30 & $1.3 10^{-2}$ \\
\hline
C1  & 40 & $0.001 $  & WR2  &479 & 0.5 & 0.63  & 0.96   & 25.71  & 21.23  & 183.97 & 8.94  & 12.92 & $3.3 10^{-3}$\\
C2  & 60 & $0.001 $  & WR2  &522 & 0.5 & 0.67  & 0.96   & 34.59  & 30.18  & 378.32 & 10.12 & 9.88  & $6.7 10^{-4}$\\
\hline
\end{tabularx}
\end{center}
$^*$ sequence ends at core hydrogen exhaustion\hfill
\end{table*}

\begin{figure}
\center
\resizebox{0.9\hsize}{!}{\includegraphics{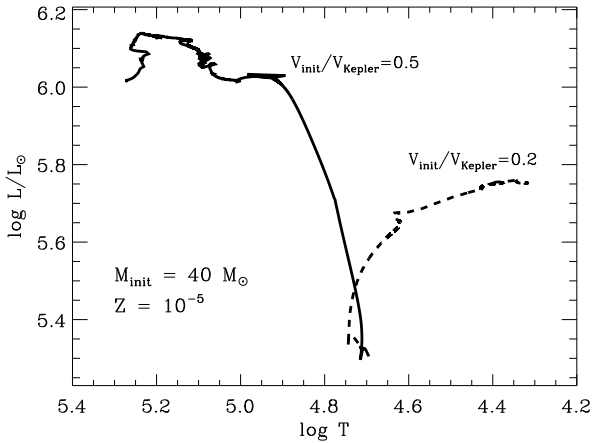}}
\caption{Evolutionary tracks for models of Seq.~3 (dashed line) and Seq.~4 (solid line), in the HR diagram (cf. Table~1), from the zero age main sequence to core carbon exhaustion. 
}\label{fig:hr}
\end{figure}

\begin{figure}
\center
\resizebox{\hsize}{!}{\includegraphics{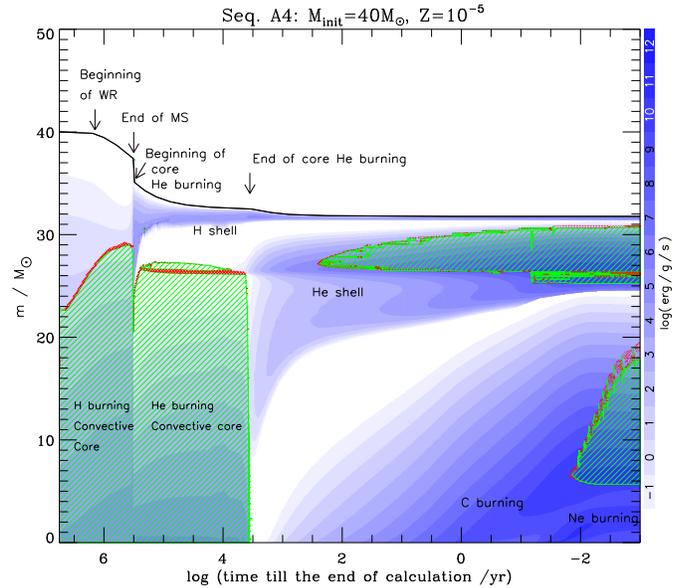}}
\caption{Evolution of the internal structure of Seq.~A4 from zero-age main sequence to
the neon burning. Convective layers are hatched. Semi-convective layers are marked
by dots. The gray shading gives nuclear energy generation rates in log scale, as indicated
on the right side. The topmost solid line denotes the surface of the star. 
}\label{fig:kipp}
\end{figure}
\begin{figure}
\center
\resizebox{\hsize}{!}{\includegraphics{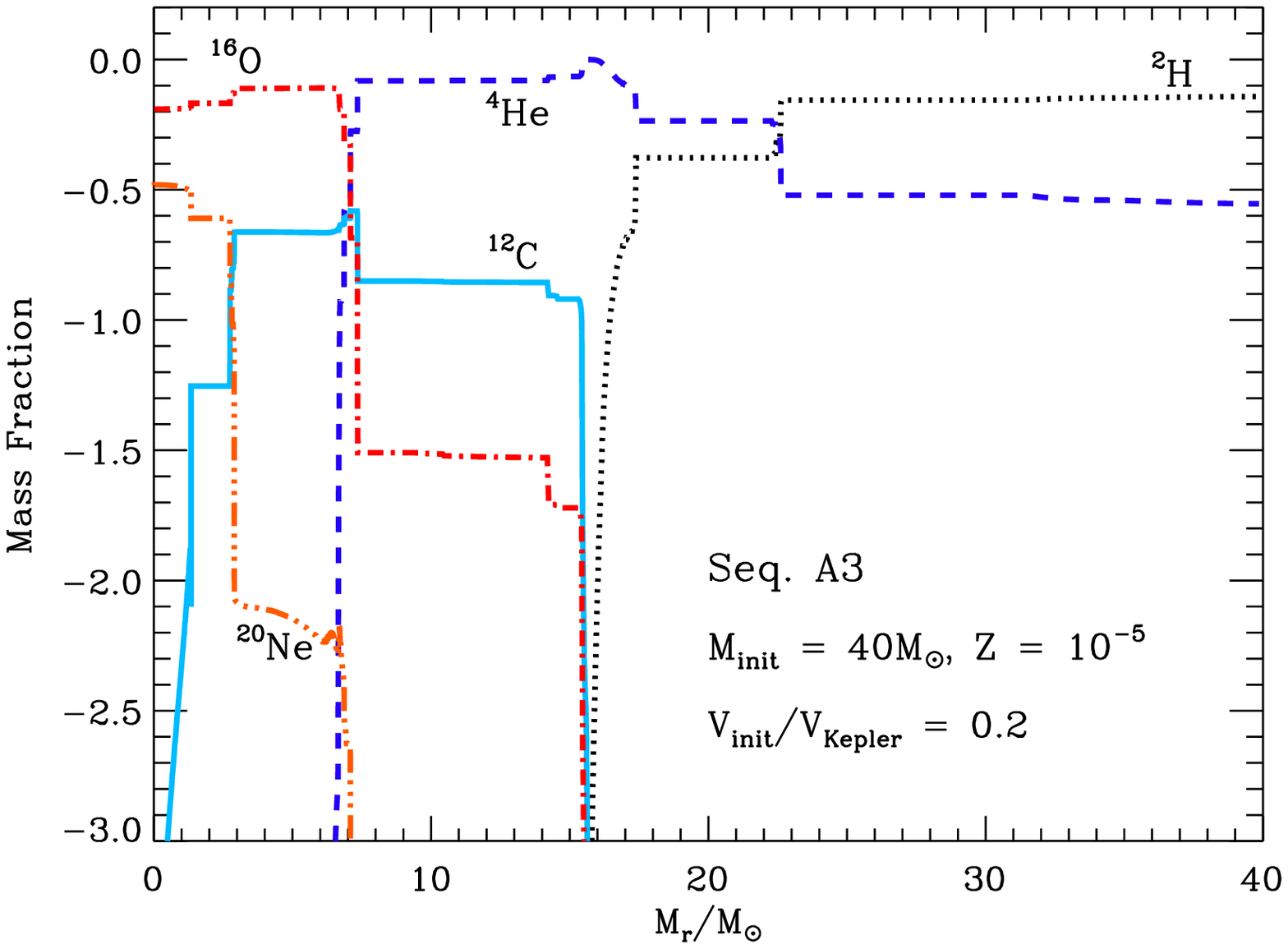}}
\resizebox{\hsize}{!}{\includegraphics{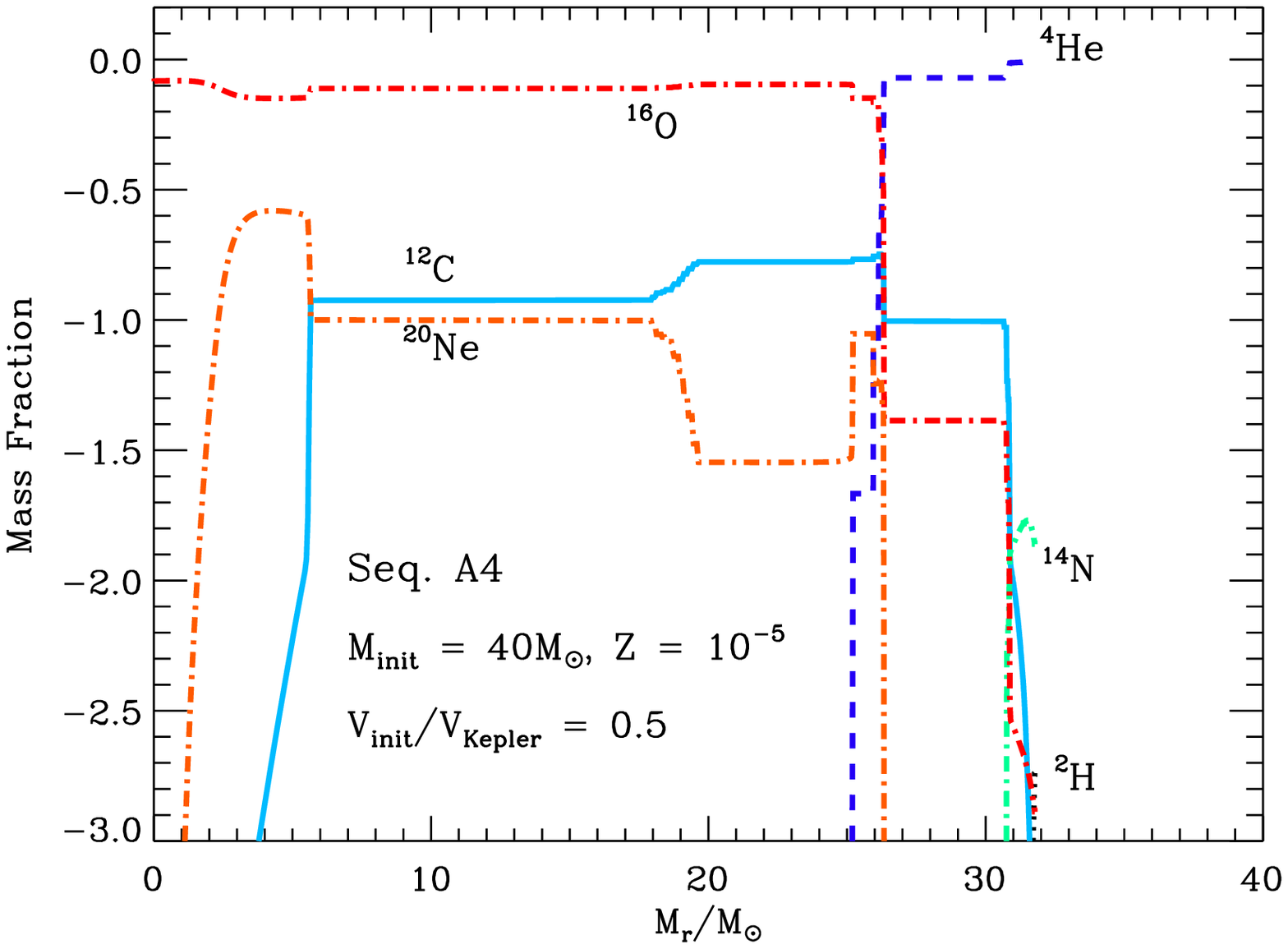}}
\caption{\emph{upper panel}: Mass fraction of chemical elements  
after core carbon exhaustion in Seq.~A3, as a function of  
the mass coordinate.
\emph{lower panel}: Same as in the upper panel, but for model Seq.~A4.
}\label{fig:chem}
\end{figure}

Table~\ref{tab1} gives an overview of the computed model sequences and their
basic properties. All model sequences with $Z=10^{-5}$ and $Z=0.001$ are followed until
the exhaustion of central carbon or the beginning of neon burning, 
after which the evolutionary times are too short ($\lsim 1$ yr)
to meaningfully change the quantities given in Table~\ref{tab1} until collapse.
The calculation for the model sequence B2 ($Z=0.02$) is stopped at the end of the main sequence. 

Fig.~\ref{fig:hr} shows that models of Seq.~A3 ($M=40~\mathrm{M_\odot}$, $f_\mathrm{K} = 0.2$) 
evolve in a similar way to non-rotating models 
of $40~\mathrm{M_\odot}$ and $Z=10^{-5}$ by Meynet \& Maeder (\cite{Meynet02}), rather than
their  rotating models which evolve further redwards.
This is because in our models, the degree of differential rotation 
is much weaker due to magnetic torque.
In Seq.~A3, helium enrichment at the end of main sequence ($\Delta Y = 0.03$) is somewhat larger
than in the 40~\Msun{} rotating model with $V_\mathrm{init} = 300~\mathrm{km~s^{-1}}$ 
of Meynet \& Maeder ($\Delta Y = 0.02$).
Recent solar metallicity models with magnetic fields 
by Maeder \& Meynet (\cite{Maeder05}) show larger
helium enrichment than their non-magnetic rotating models.
Note also that our adopted initial helium abundance is slightly higher ($Y=0.24$) 
than theirs ($Y=0.23$). 

On the other hand, models of Seq.~A4 -- which initially rotate 2 times faster than in Seq.~A3 --  
evolve bluewards due to the strong helium enrichment of the envelope
via rotationally induced chemical mixing.
The evolution of the internal structure of the star is shown
in Fig.~\ref{fig:kipp}.
The stellar mass decreases rapidly when the star enters the WR stage. 
At core hydrogen and core helium exhaustion, the stellar contraction 
leads the stellar surface to reach the $\Omega$ limit, inducing
strong centrifugally supported mass loss.
In the end, the CO core is nearly four times more massive than in Seq.~A3,
even though the star in Seq.~A4 loses more mass.
Fig.~\ref{fig:chem} compares the chemical profiles of Seq.~A3 and Seq.~A4
at the central carbon exhaustion.
In Seq.~A4, only a small amount of hydrogen is left at the stellar surface, 
but significant amounts of primary nitrogen ($X_\mathrm{^{14}N} \simeq 10^{-2}$) are present. 
In Seq.~A3, the nitrogen abundance at the surface is about $4 10^{-6}$ at this stage.

\begin{figure}
\center
\resizebox{\hsize}{!}{\includegraphics{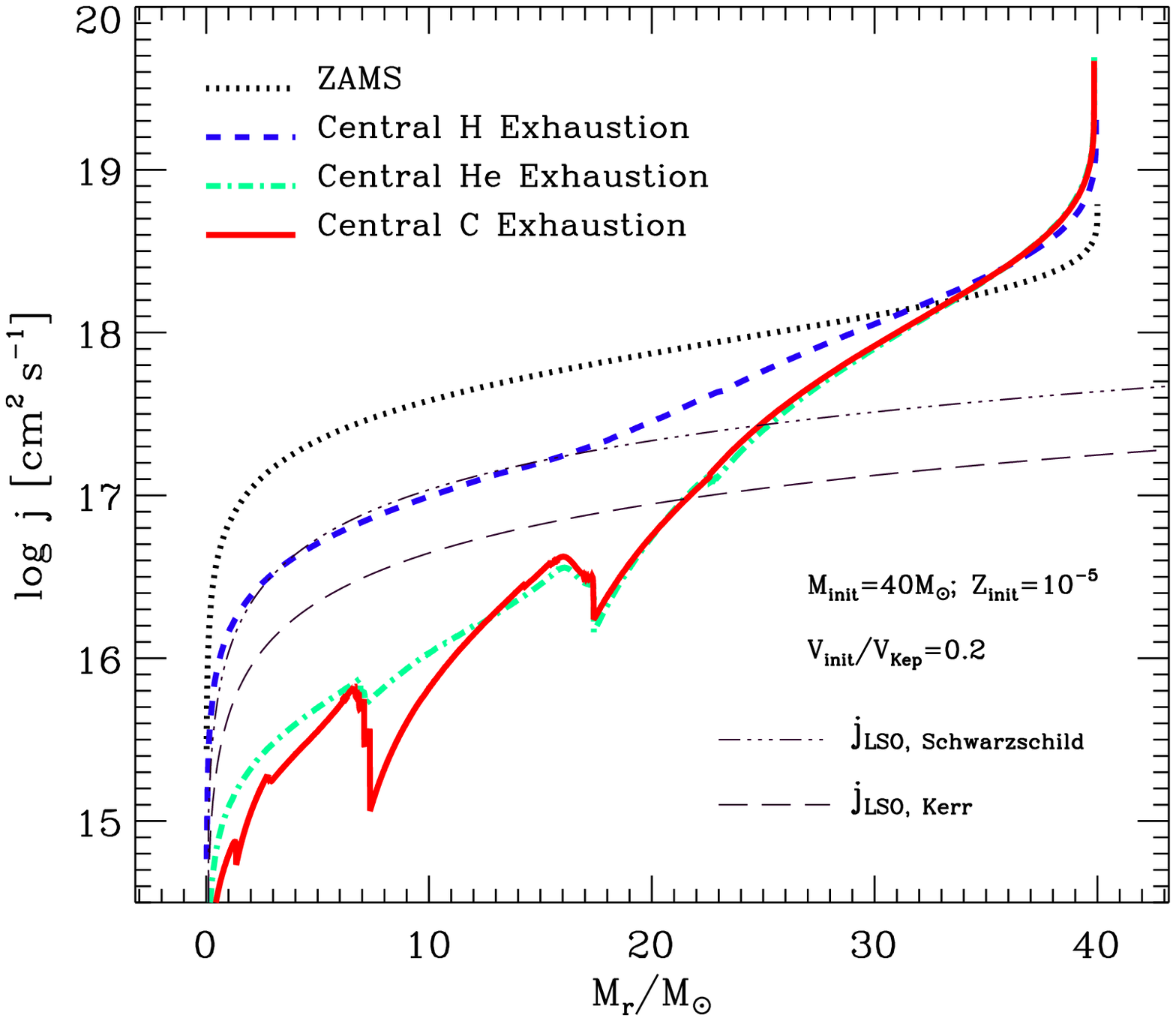}}
\resizebox{\hsize}{!}{\includegraphics{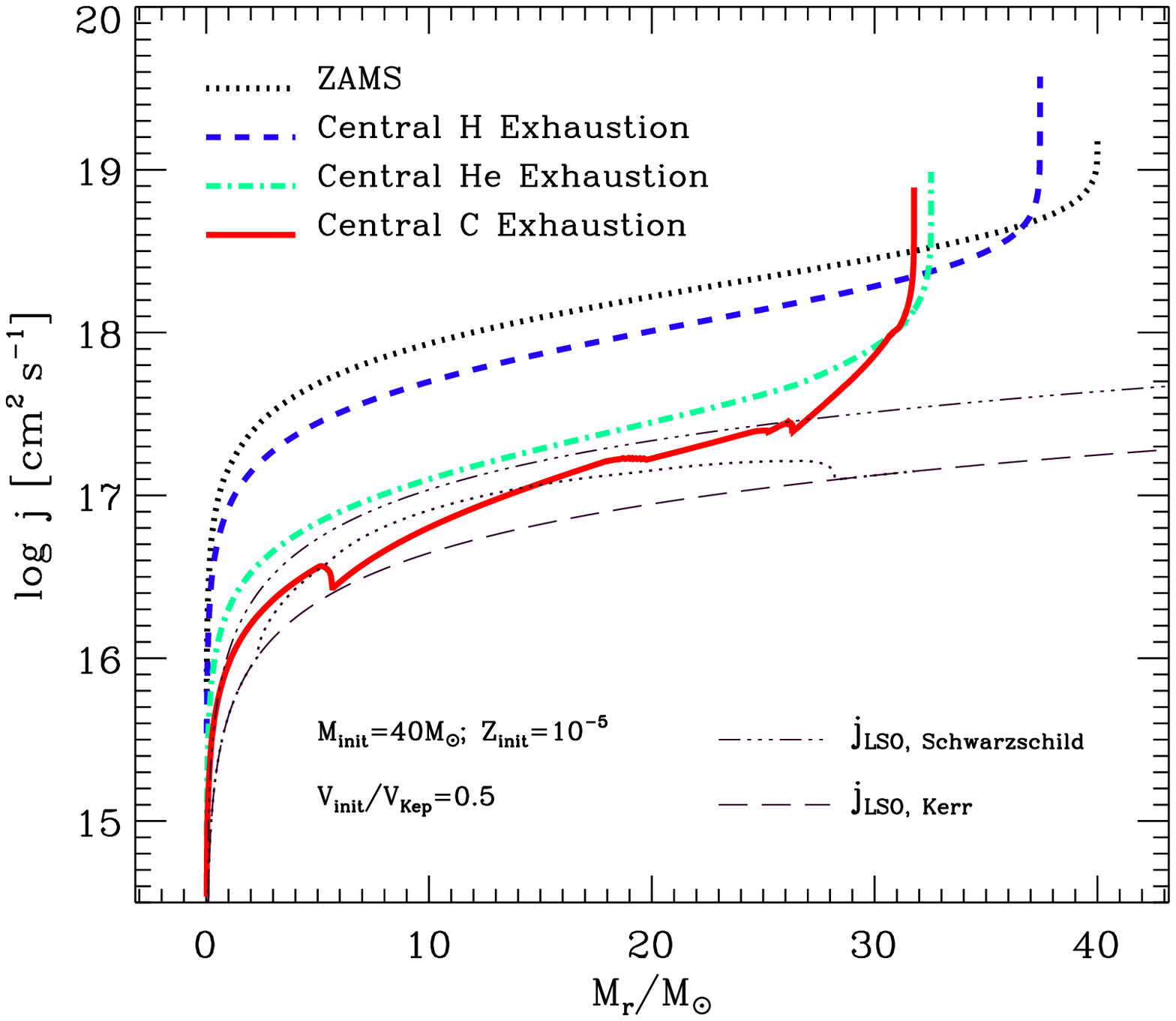}}
\caption{Mean specific angular momentum over the shells as a function of the mass coordinate
at 4 different epochs in Seq.~A3 (upper panel) and Seq.~A4 (lower panel). 
Note that the equatorial specific angular momentum is larger by a factor of 1.5.
The thin two-dotted-dashed line
denotes the angular momentum of the last stable orbit around a non-rotating Schwardzschild black hole
of mass equal to the mass coordinate. The thin long-dashed line gives the same but for a maximally rotating Kerr black hole.
The thin dotted line denotes the specific angular momentum for the last stable orbit
at the given mass of a black hole, assuming that all mass below forms a rotating black hole. 
Here, if the contained angular momentum is larger than that of a maximally rotating
black hole, the black hole is assumed to rotate maximally.
}\label{fig:jsp}
\end{figure}

\begin{figure}
\center
\resizebox{0.8\hsize}{!}{\includegraphics{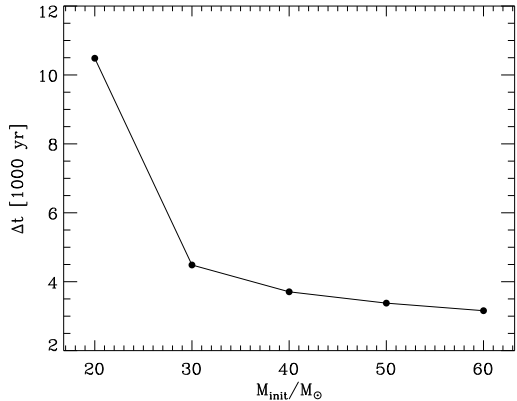}}
\caption{
Elapsed time from core helium exhaustion to the core carbon exhaustion, 
for Seqs. A1 (20~\Msun), A2 (30~\Msun), A4 (40~\Msun), A5 (50~\Msun), and A6 (60~\Msun). 
}\label{fig:ctime}
\end{figure}

We note that in Seq.~B2 ($Z=0.02$ and $f_\mathrm{K} = 0.5$), the helium enrichment
is less than in the corresponding low metallicity models (see Table~\ref{tab1}). 
The reason is that chemical mixing becomes inefficient as the star
is significantly spun down via strong stellar winds, 
given that the mixing time scale is inversely proportional to the stellar mass and spin rate
(e.g. Maeder \& Meynet~\cite{Maeder00}). 
This results indicate that homogeneous evolution is favored at lower metallicity,
since the stars suffer less mass loss.

The chemical mixing affects the internal transport of angular momentum in our models.  
Fig.~\ref{fig:jsp} shows that much more angular momentum is removed from the core
in Seq.~A3 than in Seq.~A4. 
In Seq.~A3, while stronger chemical gradient between the core
and the hydrogen envelope tends to render stronger differential rotation, 
coupling of the core with the envelope by magnetic torques enforces
efficient transport of angular momentum from the core.
In Seq.~A4, on the other hand, 
a decrease of the core angular momentum 
occurs mainly due to the stellar wind mass loss.
In this sequence, significant differential rotation only appears during the carbon core contraction.
However, carbon burning starts soon after the central helium exhaustion (about 4000 yr after), 
and a large amount of angular momentum is still preserved in the core by the central carbon exhaustion.

\begin{figure}
\center
\resizebox{0.9\hsize}{!}{\includegraphics{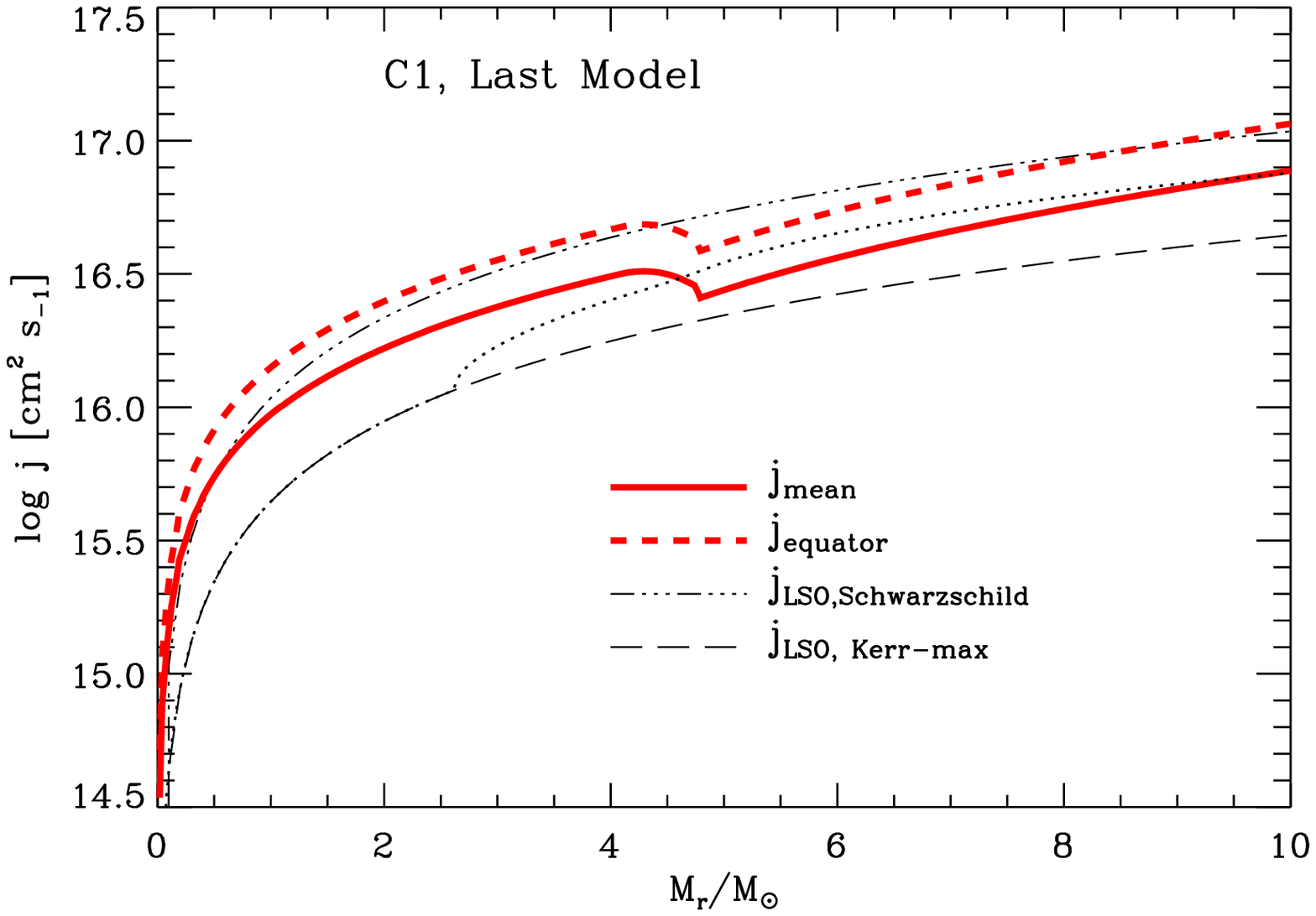}}
\resizebox{0.9\hsize}{!}{\includegraphics{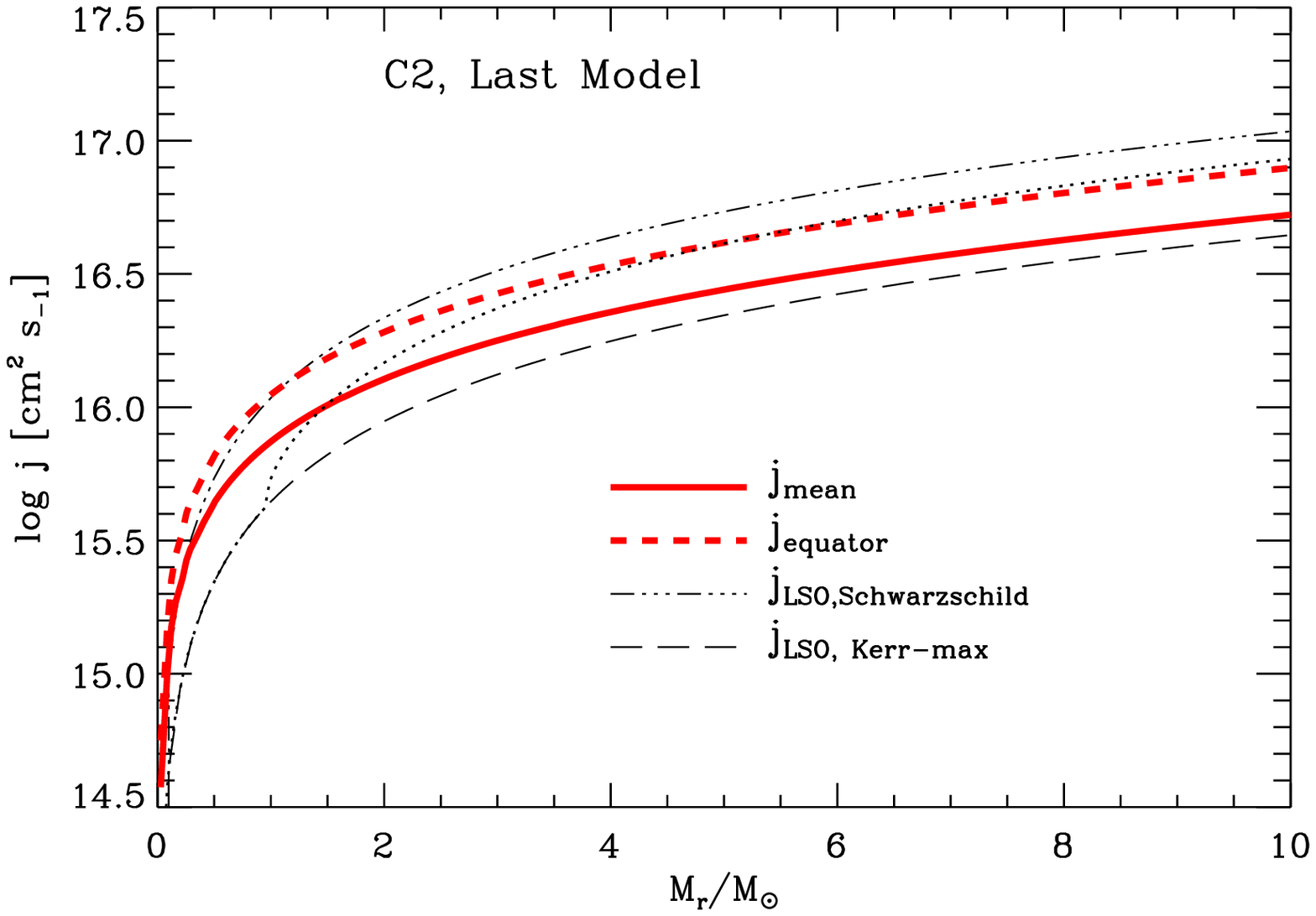}}
\caption{Specific angular momentum as a function of the mass coordinate,
in the last models of Seqs. C1 (upper panel) and C2 (lower panel) 
[ after core carbon exhaustion in Seq.~C1 and near the
core carbon exhaustion ($X_\mathrm{C} = 10^{-4}$) in Seq.~C2]. 
The thick solid line denotes the mean specific angular momentum over the shells, 
as function of the mass coordinate.
The thick dashed line gives the equatorial specific angular momentum.
The thin two-dotted-dashed line
denotes the angular momentum of the last stable orbit around a non-rotating Schwardzschild black hole
of mass equal to the mass coordinate. The thin long-dashed line gives the same but for a maximally rotating Kerr black hole.
The thin dotted line denotes the specific angular momentum of the last stable orbit
at the given mass of a black hole, assuming that all mass below forms a rotating black hole. 
Here, if the contained angular momentum is larger than that of a maximally rotating
black hole, the black hole is assumed to rotate maximally.
}\label{fig:cjsp}
\end{figure}

A comparison of the core angular momentum with the angular momentum of the last stable orbit 
around a black hole in Fig.~\ref{fig:jsp}
reveals that the Seq.~A4 fulfills the angular momentum constraints of the collapsar model. 
Table~1 indicates that indeed Sequences~A2, A4, A5 and A6, as well as C1 and C2
conserve enough core angular momentum to produce a collapsar.

On the other hand, the core of the $20~\mathrm{M_\odot}$ model of Seq.~A2
loses more angular momentum than the comparable more massive models (see Table~\ref{tab1}),
and does not retain enough angular momentum to make a collapsar.
This is mainly due to the rather long time scale of CO core contraction (Fig.~\ref{fig:ctime})
and the comparably large envelope-to-core mass ratio in this case.
This implies the possible existence of a lower mass limit for collapsar formation
through the considered evolutionary channel, which may restrict collapsars to
CO cores more massive than $10\mso$.

The CO core mass of model Seq.~A6 is about 40~\Msun{}, which roughly coincides with 
the lower mass limit of pair instability supernovae
(e.g. El~Eid \& Langer~\cite{Eid86}, Kippenhahn \& Weigert~\cite{Kippenhahn90}).
Except in a narrow mass range close to the mass limit, where pulsational
pair instability may induce a large amount of mass loss,
no stable iron cores are formed in the pair instability regime.
Black holes may still form for very high masses (cf. Heger et al.~\cite{Heger03}).
However, for a normal initial mass function, CO cores of about $40\mso$ may be
considered as upper mass limits for collapsar production via chemically
homogeneous evolution.

In Seq.~B2, where metallicity is hundred times higher ($Z=0.001$) than in Seq.~A4, 
quasi-chemically homogeneous evolution still occurs, but the mass loss in this $40\mso$ sequence 
spins down the core to angular momentum values below the collapsar requirement.
Sequences~C2 and C3, however, show that with the currently best estimate for helium star 
winds (Vink \& de Koter~\cite{Vink05}), collapsars are still being produced at $Z=0.001$
(Table~\ref{tab1}; Fig.~\ref{fig:cjsp}).
Given the still significant remaining uncertainties in Wolf-Rayet wind models,
which partly imply very large clumping factors and thus further reductions in the
Wolf-Rayet mass loss rates (Graefener \& Hamann~\cite{Graefener05}), 
it appears conceivable that the considered evolutionary channel
might be relevant for GRBs not only at high redshifts, but also in
the local universe.

\section{Discussion}\label{sect:discussion}

Here, we investigate the evolution of rotating massive single stars at low metallicity,
with the Tayler-Spruit dynamo included. 
Our results show that if the initial spin rate is high enough,  such that
the time scale for rotationally induced mixing becomes shorter than the nuclear time
(Maeder \& Meynet~\cite{Maeder00}),
the massive star may evolve in a quasi-chemically homogeneous way to become
a rapidly rotating WR star on the main sequence (Maeder~\cite{Maeder87}; Langer~\cite{Langer92}). 
In particular, it is shown that for low enough metallicity, this type of evolution 
can lead to retention of sufficient 
angular momentum in CO cores in the range $10\mso \dots 40\mso$
to produce gamma-ray bursts according to the 
collapsar scenario (Woosley~\cite{Woosley93}).

The critical values of metallicity and initial spin beyond which collapsars can be
produced depend on uncertainties in the description of rotationally induced
mixing and stellar wind mass loss. 
The Wolf-Rayet wind mass loss rate may be of crucial importance, as illustrated above.
In general, more collapsars are produced through chemically homogeneous evolution for lower 
metallicity, since less mass loss by stellar winds will keep the star
more rapidly rotating, which favors mixing. 

However, due to the uncertainties in the stellar wind mass loss rates, and
the uncertain distribution of initial stellar rotation rates (Maeder \& Eenens\cite{Maeder04}),
chemically homogeneous evolution at near solar metallicity can not be excluded.
Observations actually  provide evidence for homogeneous evolution
in some young star clusters in the Galaxy and the Magellanic Clouds 
(e.g. Howarth \& Smith~\cite{Howarth01}; Bouret et al.~\cite{Bouret03};
Walborn et al.~\cite{Walborn04};).
Although such findings are infrequent, this may in fact be consistent
with the observationally implied low frequency of GRBs.
Importantly, the results of Seq.~C1 and C2 with $Z=0.001$ indicate that
the blue stars as found by Bouret et al.~(\cite{Bouret03})
and Walborn et al.~(\cite{Walborn04}) in the Magellanic Clouds
which have the signature of homogeneous evolution
might be promising progenitors of GRBs.
Our solar metallicity Sequence~B2 shows, on the other hand, the quasi-chemically homogeneous evolution
does not always lead to a collapsar, in particular at high metallicity.

It appears likely that at sufficiently low metallicity, 
the chemically homogeneous evolution channel 
will dominate over binary collapsar production channels
as the helium star merger scenario or the tidal spin-up of helium stars ---
which are considered as a major path towards collapsars
(e.g. Izzard et al.~\cite{Izzard04}; Podsiadlowski et al.~\cite{Podsiadlowski04};
Fryer \& Heger~\cite{Fryer05}): 
very metal poor stars may not evolve into red super-giants, and common envelope
evolution without merging of the two stars, which is required in both mentioned scenarios,
may be difficult.
At zero metallicity, it is even questioned whether binary stars can form at all
(Bromm \& Larson~\cite{Bromm04}).
Therefore, if GRBs occur from the first generations of stars,
chemically homogeneous evolution may provide their major formation path.

\begin{acknowledgements}
We are grateful to Alexander Heger \& Arend-Jan Plarends for technical assistance.
We would like to thank Alex de Koter for the fruitful discussion about WR winds.
Sung-Chul Yoon acknowledges the support by the Netherlands Organization
for Scientific Research (NWO) through the VENI grant 
(639.041.406). 
\end{acknowledgements}


\end{document}